\documentclass[aps,prl,floatfix,showpacs,twocolumn]{revtex4}
\usepackage{amsmath,amssymb}
\usepackage{epsfig}

\newcommand{\bm}{\boldsymbol}

\begin{document}

\hsize\textwidth\columnwidth\hsize\csname@twocolumnfalse\endcsname

\title{Chirality-induced Dynamic Kohn Anomalies in Graphene}

\author{Wang-Kong Tse$^1$}
\author{Ben Yu-Kuang Hu$^{1,2}$}
\author{S. Das Sarma$^1$}
\affiliation{$^1$Condensed Matter Theory Center, Department of Physics,
University of Maryland, College Park, Maryland 20742}
\affiliation{$^2$Department of Physics, University of Akron, Akron, Ohio 44325-4001}

\begin{abstract}
We develop a theory for the renormalization of the phonon energy dispersion in graphene due to the combined effects of both Coulomb and electron-phonon (e-ph) interactions. 
We obtain the renormalized phonon energy spectrum by an exact analytic derivation of the phonon self-energy, finding three distinct Kohn anomalies (KAs) at the phonon wavevector $q = \omega/v, 2k_F\pm\omega/v$ for LO phonons and one at $q = \omega/v$ for TO phonons. The presence of these new KAs in graphene, in contrast to the usual KA $q = 2k_F$ in ordinary metals, originates from the dynamical screening of e-ph interaction (with a concomitant breakdown of the Born-Oppenheimer approximation) and the peculiar chirality of the graphene e-ph coupling. 
\end{abstract}


\maketitle

Graphene, a two-dimensional (2D) sheet of graphite comprising a planar hexagonal lattice of carbon atoms, has attracted a great deal of recent interest with enormous amount of experimental and theoretical activity since its discovery. Of fundamental significance is the study of the interaction between electrons and phonons in this novel system. In particular, the many-body renormalization of the phonon spectrum due to the electron-phonon (e-ph) interaction is an important open question. Experimentally, Raman scattering offers a means to measure the long-wavelength phonon energy whereas X-ray and neutron scattering provide an avenue to probe the entire phonon energy dispersion. Due to free carrier-induced many-body interactions, one expects the observed phonon energy in a doped system to be different from the ``bare'' phonon energy in an undoped system. Physically, electrons respond to the dynamical lattice vibrations by screening the lattice potential, changing the elastic constants of these vibrational modes and thereby renormalizing the phonon energy. Recently, a number of Raman scattering experiments \cite{Raman} on extrinsic graphene (where the electron density can be tuned by gating) 
have emerged, with the observed density dependence of the Raman shift (for the long-wavelength G-band optical phonons at the $\Gamma$ point) pointing to the inapplicability of the Born-Oppenheimer approximation (BA) \cite{Rem1} in graphene. This density dependence has been addressed for the long-wavelength phonons (i.e. $q = 0$) using perturbation theory \cite{Ando1,Raman} and density-functional theory \cite{Mauri}.

An interesting question remains as to whether the Kohn anomaly (KA) \cite{KA}, which appears as a cusp in the phonon energy dispersion at $q = 2k_F$ for ordinary 2D metals, will be substantially modified in graphene due to its quasi-relativistic chiral band structure. The occurrence of KA is entirely a many-body effect as it originates from the screening of the e-ph interaction by electrons, and correpsonds to the singularities of the phonon self-energy or its derivatives as a function of $q$. 
For graphite, the KA was recently studied in Ref.~\cite{Piscanec} as a function of $q$ at zero doping. Ref.~\cite{Mauri} addresses the phonon energy renormalization in doped graphene, but only at $q = 0$ relevant to Raman scattering experiments. Traditionally, KA is probed experimentally with X-ray or neutron scattering spectroscopy by measuring the phonon energy as a function of $q$, for which a systematic theory for the phonon energy renormalization in graphene for finite wavevector $q > 0$ is still lacking. 

In this Letter, we present a theory for the renormalization of the phonon energy dispersion in graphene due to both Coulomb and e-ph interaction effects. 
%
%
From our theory, we obtain two major new results: (1) We find that direct Coulomb and phonon-mediated electron-electron interactions decouple to all orders of perturbation theory within the random-phase approximation (RPA), and the electronic collective plasmon mode does not contribute to the phonon energy renormalization; 
(2) we obtain the renormalized phonon energy dispersion as a function of $q$, predicting the occurrence of three distinct KAs at the phonon wavevector $q = \omega/v, 2k_F\pm\omega/v$ for the LO mode and one at $q = \omega/v$ for the TO mode, which arise from the chiral structure of the graphene e-ph coupling. The novel pecularity that these KAs do not occur at $q = 2k_F$ (as in usual metals) originates from the fact that the phonon dynamics cannot be neglected in the screened e-ph interaction, indicating the inapplicability of the BA.

Graphene, behaving as a 2D zero-gap semiconductor, has a G-band optical phonon energy $\omega_0 = 200\textrm{meV}$ at the $\Gamma$ point which is comparable to the Fermi energy $\varepsilon_F \sim 110-370\textrm{meV}$ at the usual extrinsic carrier density $n = 10^{12}-10^{13}\textrm{cm}^{-2}$. This approximate equality of the phonon and electron energy scales $\omega_0 \sim \varepsilon_F$ implies a breakdown of the static approximation for the phonon degree of freedom \cite{Rem2}, naturally explaining the violation of the BA in the recent Raman scattering experiments \cite{Raman} since the BA implicitly assumes the phonon dynamics to be much slower than the electron dynamics.  In addition, plasmon-phonon coupling, which has been extensively studied in doped semiconductor systems (e.g., GaAs, SiC), is expected to occur whenever the phonon dynamics is at a comparable time scale as the electron motion. It follows that one cannot take the screening of the e-ph interaction to be simply static while keeping the screening of the Coulomb interaction to be dynamic; instead, one has to take into account the dynamical screening of both Coulomb and e-ph interactions, i.e., direct electron-electron (e-e) and e-ph interactions must be treated on an equal footing \cite{Jala}.

Near the Brillouin zone corner K (i.e. the Dirac point), graphene is described by the effective chiral Hamiltonian $H = v\bm{\sigma}\cdot\bm{k}$, where $v \approx 10^6\textrm{ms}^{-1}$ is the quasiparticle velocity and $\bm{\sigma}$ is the set of Pauli matrices describing the two $A$ and $B$ sublattice degrees of freedom. The quasiparticle energy dispersion as obtained from this Hamiltonian is $\epsilon_{k\lambda} = \lambda\epsilon_k$, where $\epsilon_k = vk$ and $\lambda$ is the chirality label representing the conduction band ($\lambda = 1$) and valence band ($\lambda = -1$); the corresponding eigenstate is denoted as $\vert\bm{k}\lambda\rangle$. 
The e-ph interaction vertex for optical phonons is given by \cite{Ando,Ando1,phonon} $g\bm{M}$, where $g$ is the coupling constant characterizing the magnitude of the e-ph interaction, and $\bm{M}$ is the off-diagonal matrix 
\begin{eqnarray}
\bm{M}(\bm{q}) = \left[\begin{array}{cc}
0 & M_{{\mathrm{AB}}}e^{-i\phi_q} \\
M_{{\mathrm{BA}}}e^{i\phi_q} & 0
\end{array}\right],
\label{eq1}
\end{eqnarray}
with $M_{\mathrm{AB}} = -1$ or $i$ and $M_{\mathrm{BA}} = 1$ or $i$ for LO or TO phonons, respectively, and $\phi_q = \mathrm{tan}^{-1}(q_y/q_x)$ the azithmuthal angle of the momentum $\bm{q}$. The chiral structure of the e-ph coupling is peculiar to graphene, describing the bond stretching and bending between neighbouring carbon atoms of the A-sublattice and B-sublattice. 

\begin{figure}[h]
  \includegraphics[width=8.0cm,angle=0]{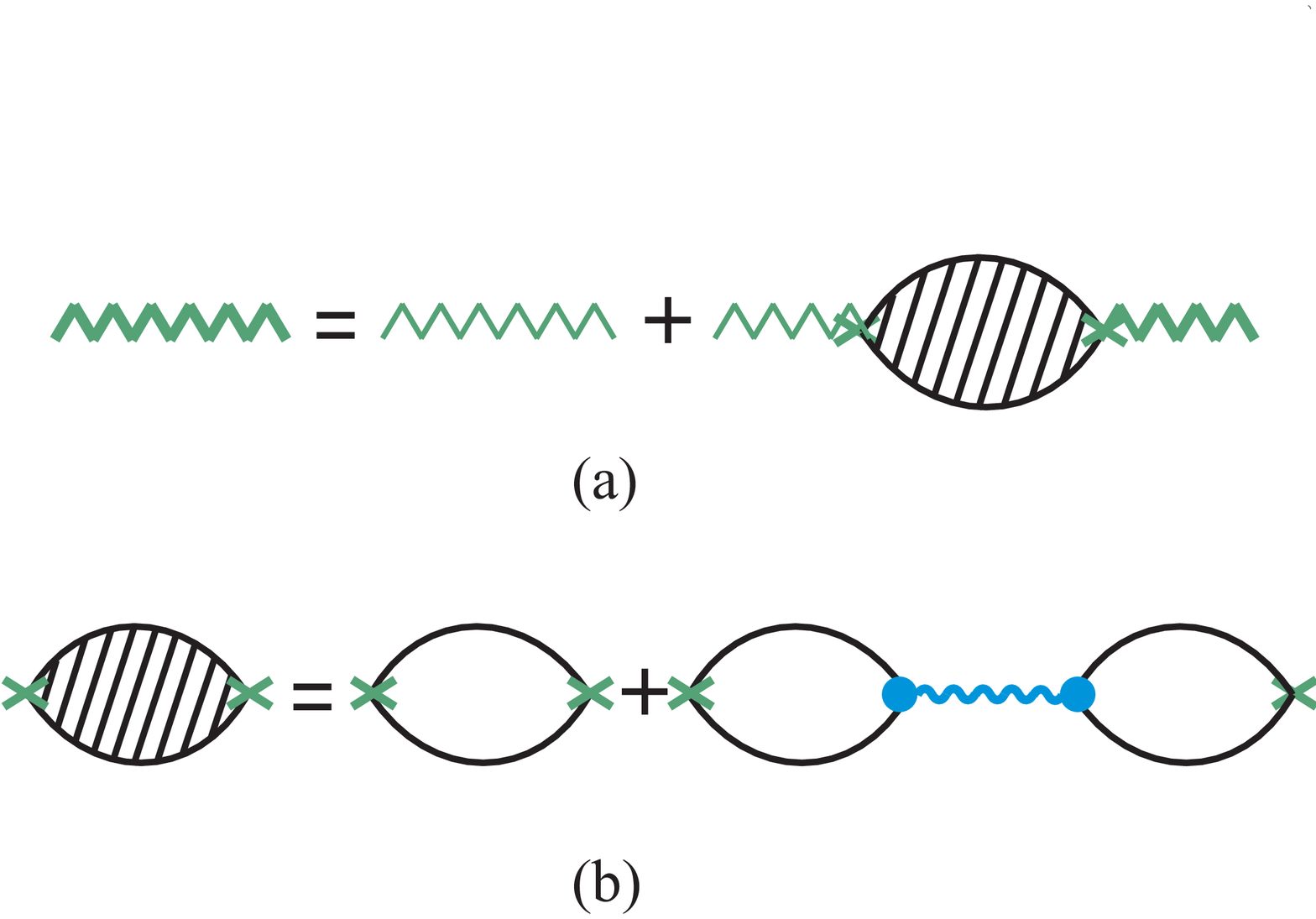}  
\caption{(Color online) (a) Dyson equation for the renormalization of the phonon Green function. The zigzag lines denote the phonon Green function and the crosses denote the e-ph interaction vertex. The shaded bubble with two cross vertices stands for the renormalized phonon self-energy. (b) Equation for the renormalized phonon self-energy. The unshaded bubble with two cross vertices denotes the bare phonon self-energy, the two bubbles with one cross vertex and one dot vertex are ``hybrid bubbles'' with one e-ph interaction vertex and one Coulomb interaction vertex. The wavy line stands for the usual RPA-screened Coulomb interaction.} \label{fig5}
\end{figure}

In the presence of direct e-e interaction and e-ph interaction, the phonon Green function is renormalized by both, as represented in the diagrammatic language in Fig.~\ref{fig5}.
Summation of these diagrams leads to the renormalized phonon Green function $D(q,\omega) = {D^0(q,\omega)}/[{1-D^0(q,\omega)\Pi^{\mathrm{pp}}(q,\omega)}]$, 
%
%
where $D^0(q,\omega) = 2\omega_0/(\omega^2-\omega_0^2+i0^+)$ is the bare phonon Green function with $\omega_0$ the optical phonon energy, $\Pi^{\mathrm{pp}}(q,\omega)$ is the RPA-screened phonon self-energy given by the following equation [Fig.~\ref{fig5}(b)]:
\begin{equation}
\Pi^{\mathrm{pp}}(q,\omega) = \Pi_0^{\mathrm{pp}}(q,\omega)+\Pi_0^{\mathrm{pc}}(q,\omega)V_{\mathrm{ee}}^{\mathrm{c}}(q,\omega)\Pi_0^{\mathrm{cp}}(q,\omega),
\label{eq2}
\end{equation}
where $V_{\mathrm{ee}}^{\mathrm{c}}(q,\omega) = V_q/[1-V_q\Pi_0^{\mathrm{cc}}(q,\omega)]$ is the usual RPA-screened Coulomb interaction \cite{gpFL} (here $\Pi_0^{\mathrm{cc}}$ is the electronic polarizability), $\Pi_0^{\mathrm{pp}}(q,\omega)$ the bare phonon self-energy, and $\Pi_0^{\mathrm{pc}}(q,\omega)$, $\Pi_0^{\mathrm{cp}}(q,\omega)$ are the bare ``hybrid'' bubbles with one Coulomb interaction vertex and one e-ph interaction vertex. 
In regular metals or doped polar semiconductors, the phonon self-energy and the hydrid bubbles are the same (up to a factor given by the e-ph coupling constant) as the electronic polarizability; the Dyson equation for the renormalized phonon Green function (Fig.~\ref{fig5}) simply reduces to the usual RPA series. In graphene, due to the presence of a chiral structure of the e-ph interaction Eq.~(\ref{eq1}),
the bubbles $\Pi_0^{\textrm{pp}}$, $\Pi_0^{\textrm{pc}}$, $\Pi_0^{\textrm{cp}}$, and $\Pi_0^{\textrm{cc}}$ are not equal to one another. In particular, $\Pi_0^{\textrm{pc}}$ and $\Pi_0^{\textrm{cp}}$ incorporate the coupling effects of the Coulomb and e-ph interactions within the RPA, and describe the renormalization effect of the phonon energy by the Coulomb interaction. 
Interestingly, we find that with the graphene e-ph interaction Eq.~(\ref{eq1}), these hybrid bubbles vanish identically: 
$\Pi^{\mathrm{pc}}_0 = \Pi^{\mathrm{cp}}_0 =k_{\mathrm{B}}Tg\sum_{ik_n,k}G^0_{k\lambda}(ik_n)G^0_{k+q\lambda'}(ik_n+iq_n)\langle\bf{k}+\bf{q}\lambda'\vert M(\bf{q}) \vert \bf{k}\lambda \rangle \langle \bf{k}\lambda \vert \bf{k}+\bf{q}\lambda'\rangle = 0$,
%
%
where $G^0_{k\lambda}(ik_n) = 1/(ik_n-\xi_{k\lambda})$ is the quasiparticle Green function with $\xi_{k\lambda} = \epsilon_{k\lambda}-\varepsilon_F$ the quasiparticle energy rendered from the Fermi level. Eq.~(\ref{eq2}) then implies that $\Pi^{\mathrm{pp}}$ is simply given by the bare phonon self-energy $\Pi_0^{\mathrm{pp}}$, and Coulomb interaction does not contribute to the screening of the e-ph interaction within the RPA. The phonon energy dispersion, which is given by the pole of the real part of the renormalized phonon Green function, 
\begin{equation}
\omega^2 = \omega_0^2+2\omega_0\mathrm{Re}\Pi_0^{\mathrm{pp}}(q,\omega),
\label{eq4}
\end{equation}
is therefore only renormalized by the e-ph interaction but not by Coulomb interaction. It follows that coupled plasmon-phonon modes do not arise in graphene, with the phonon and plasmon modes having separate branches of energy dispersion in the $\omega-q$ phase space despite comparable energy scales for the phonon and electron dynamics. In addition, direct Coulomb and phonon-mediated e-e interactions are simply additive, with the RPA-screened total e-e interaction given by $V_{\textrm{ee}}^{\textrm{tot}} = V_q/[1-V_q\Pi_0^{\mathrm{cc}}]+V_{\textrm{ee}}^{\textrm{ph}}/[1-D^0\Pi_0^{\textrm{pp}}]$, where $V_{\textrm{ee}}^{\textrm{ph}}$ is the unscreened phonon-mediated e-e interaction \cite{phonon}. 
%
\begin{figure}[h]
  \includegraphics[width=7.5cm,angle=0]{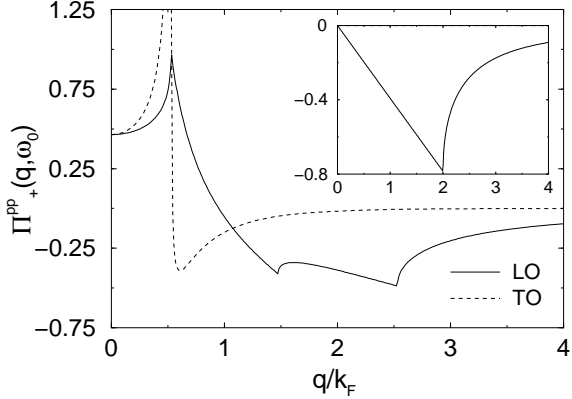} 
\caption{Dynamic phonon self-energy $\tilde{\Pi}^{\mathrm{pp}}_+(x,u)$ versus $x = q/k_F$ at $u = \omega_0/\varepsilon_F$ for the LO (solid line) and TO (dashed) modes at a density $n = 10^{13}\mathrm{cm}^{-2}$.
Inset: Static phonon self-energy $\tilde{\Pi}^{\mathrm{pp}}_+(x,u)$  versus $x = q/k_F$ at $u = 0$. 
} \label{fig2}
\end{figure}

The phonon self-energy $\Pi^{\mathrm{pp}} = \Pi_0^{\mathrm{pp}}$ can then be derived from the e-ph interaction vertex Eq.~\ref{eq1} as:
%
%
\begin{eqnarray}
\Pi^{\mathrm{pp}}(q,iq_n) &=& 4\frac{1}{2}g^2
\sum_{\lambda\lambda'}\sum_{k}\frac{n_F(\xi_{k\lambda})-n_F(\xi_{k+q\lambda'})}{iq_n+\xi_{k\lambda}-\xi_{k+q\lambda'}} \nonumber \\
&&\left[1\mp\lambda\lambda'\mathrm{cos}(\phi_{k+q}+\phi_k)\right],
\label{eq5}
\end{eqnarray}
where the sign $-(+)$ corresponds to LO(TO) phonons, the factor of $4$ counts the spin and valley degeneracies, and $\phi_k = \mathrm{tan}^{-1}(k_y/k_x)$ is the azithmuthal angle of the momentum $\bf{k}$ measured from $\bf{q}$. The real part of the phonon self-energy comprises two contributions $\mathrm{Re}\Pi^{\mathrm{pp}} \equiv \Pi^{\mathrm{pp}}_+ + \Pi^{\mathrm{pp}}_-$ (dropping the notation `$\mathrm{Re}$' for the real part thereafter), where $\Pi^{\mathrm{pp}}_{\mu} = 2g^2\sum_{\lambda}\sum_{k}n_F(\xi_{k\mu})f_{\mu\lambda}(\bm{k},\bm{q})$ with $f_{\mu\lambda}(\bm{k},\bm{q}) = \left[1\mp\lambda\mathrm{cos}(\phi_{k+q}+\phi_k)\right]/\left[\omega+\mu\xi_{k+}-\xi_{k+q\lambda}\right]-\left[1\mp\lambda\mathrm{cos}(\phi_{k}+\phi_{k-q})\right]/\left[\omega-\mu\xi_{k+}+\xi_{k-q\lambda}\right]$.
%
%
%
The contribution ${\Pi}^{\mathrm{pp}}_+$ is due to the extrinsic conduction band electrons whereas ${\Pi}^{\mathrm{pp}}_-$ is solely due to the intrinsic valence band electrons. In the renormalization of the phonon energy in extrinsic graphene, ${\Pi}^{\mathrm{pp}}_-$ should be subtracted from the total ${\Pi}^{\mathrm{pp}}$ to avoid overcounting of the intrinsic contribution, since the ``bare'' phonon energy (i.e. when graphene is undoped), by definition, already includes the effect of ${\Pi}^{\mathrm{pp}}_-$ \cite{Ando,Ando1}. Therefore, only ${\Pi}^{\mathrm{pp}}_+$ should be taken into account in the phonon energy renormalization by free carriers in extrinsic graphene. We also note that, for ${\Pi}^{\mathrm{pp}}_+$, the largest energy scale is the Fermi energy $\varepsilon_F$ whereas for ${\Pi}^{\mathrm{pp}}_-$ it is much higher, of the order of the cutoff energy $\Lambda = 2\pi\hbar v/a$ for the graphene linear band dispersion ($a = 2.46\mathrm{\AA}$ is the graphene lattice spacing). Therefore, although the BA is inapplicable for extrinsic graphene with $\omega_0/\varepsilon_F \sim \mathcal{O}(1)$, for intrinsic graphene the BA is strictly valid since $\omega_0/\Lambda \ll 1$. 

For clarity, we express our results in the following dimensionless quantities: wavevector $x = q/k_F$, energy $u = \omega/\varepsilon_F$, and phonon self-energy $\tilde{\Pi}^{\mathrm{pp}} = \Pi^{\mathrm{pp}}/(2g_{\textrm{ee}}^2\varepsilon_{\textrm{F}}/\pi)$ with $g_{\textrm{ee}}^2 = g^2\mathcal{A}/\hbar^2v^2$ the dimensionless phonon-mediated e-e coupling constant ($\mathcal{A}$ is the sample area). We have obtained the following asymptotic results for the phonon self-energy in the limit $q/k_F \ll 1$: $\tilde{\Pi}^{\mathrm{pp}}_{\textrm{LO,TO}+}(x,u) = (1/2)\left[1+(u/4)\mathrm{ln}\left\vert(2-u)/(2+u)\right\vert\right]+\Delta\tilde{\Pi}^{\mathrm{pp}}_{\textrm{LO,TO}+}(x,u)$, where the next order correction to the long wavelength result is given by,
%
\begin{eqnarray}
\Delta\tilde{\Pi}^{\mathrm{pp}}_{\textrm{LO}+}(x,u) &=& \frac{1}{2}\left[\frac{8-4u^2-u^4}{2u^2(u^2-4)^2}-\frac{1}{8u}\mathrm{ln}\left\vert\frac{2-u}{2+u}\right\vert\right]x^2, \label{eq8} \\
\Delta\tilde{\Pi}^{\mathrm{pp}}_{\textrm{TO}+}(x,u) &=& \frac{1}{2}\left[\frac{24-12u^2+u^4}{2u^2(u^2-4)^2}+\frac{1}{8u}\mathrm{ln}\left\vert\frac{2-u}{2+u}\right\vert\right]x^2. \nonumber \\ \label{eq9}
\end{eqnarray}
At $x = 0$, the long wavelength result \cite{Raman,Mauri,Guinea} is the same for LO and TO phonons, the two phonon modes being degenerate at the $\Gamma$ point. At finite wavevector $x$, this degeneracy is lifted with the leading-order correction going as $x^2$ given by Eqs.~(\ref{eq8})-(\ref{eq9}). We have also obtained the following analytic results for the phonon self-energy in the static case $u = 0$: $\tilde{\Pi}^{\mathrm{pp}}_{\textrm{TO}+}(x,0) = 0$, and $\tilde{\Pi}^{\mathrm{pp}}_{\textrm{LO}+}(x,0) = -(\pi/8)x\theta(2-x)+({1}/{4})[(2/x)\sqrt{x^2-4}-x\mathrm{tan}^{-1}({2}/{\sqrt{x^2-4}})]\theta(x-2)$.
%

%
%
The above static phonon self-energy results are depicted in the inset of Fig.~\ref{fig2}, which clearly shows the presence of a non-analyticity at $q = 2k_F$ corresponding to the KA for LO phonons. We also note that this non-analyticity is entirely absent in the static electronic polarizability \cite{Polar} of graphene. This is in contrast to the situation for regular materials with a parabolic energy dispersion where the phonon self-energy and the polarizability are equal up to a factor given by the e-ph coupling. This distinctive difference between the phonon self-energy and the polarizability in graphene is a direct result of the presence of a chiral structure in the graphene e-ph coupling Eq.~(\ref{eq1}), which leads to a different Berry phase dependence in the expression of the phonon self-energy Eq.~(\ref{eq5}) compared with the polarizability \cite{Polar}. Therefore, KAs in graphene originate entirely from the special chiral structure of the e-ph coupling Eq.~(\ref{eq1}).  

We have evaluated the full expression for $\Pi_+^{\mathrm{pp}}$ with general $q$ and $\omega$ dependence analytically \cite{next}, which is however too cumbersome to be shown here. The main plot of Fig.~\ref{fig2} shows the evaluated $\Pi_+^{\mathrm{pp}}$ as a function of $x$ at the phonon energy $\omega = \omega_0$, from which three cusps occurring at $vq = \omega_0$, $vq = 2\varepsilon_F\pm \omega_0$ for the LO mode are clearly discernible. For the TO mode, there is a divergence of $\Pi_+^{\mathrm{pp}}$ at $vq = \omega_0$. These non-analyticities correspond to the values of $q$ where the denominator of the integrand of $\Pi_+^{\mathrm{pp}}$ vanishes at the Fermi surface $k = k_F$, i.e. the zeros of the equation $\omega\pm vk_F\pm v\vert k_F\pm q\vert = 0$. 
%
%

\begin{figure}[h]
  \includegraphics[width=7.8cm,angle=0]{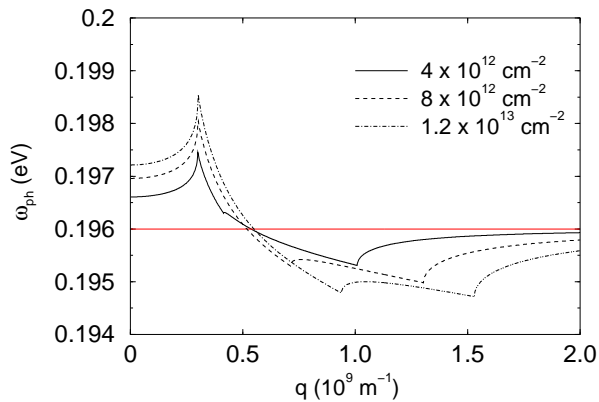} 
\caption{Renormalized LO phonon energy spectrum $\omega_{\mathrm{ph}}$ versus $q$ at different electron densities $n$. The bare phonon energy is shown as the horizontal solid line. The range of phonon wavevector $q$ shown corresponds to $[0,0.08(2\pi/a)]$ away from the $\Gamma$ point.} \label{fig3}
\end{figure}

\begin{figure}[h]
  \includegraphics[width=7.8cm,angle=0]{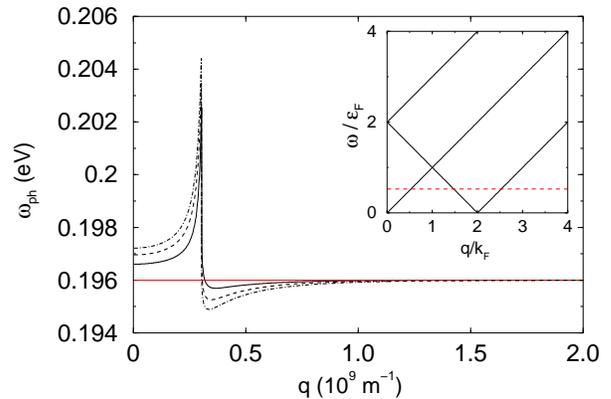}
\caption{Renormalized TO phonon energy spectrum $\omega_{\mathrm{ph}}$ versus $q$. The legends are the same as in Fig.~\ref{fig3}. Inset: Different regions for the analytical behavior of $\Pi_+^{\mathrm{pp}}$. The solid lines indicate the boundaries for these regions, and the dashed line shows the phonon energy $\omega = \omega_0$ at a density $n = 10^{13}\mathrm{cm}^{-2}$.} \label{fig4}
\end{figure}

With the calculated phonon self-energy, the renormalized phonon energy spectrum can be obtained by self-consistently solving Eq.~(\ref{eq4}) for $\omega$, which is shown in Fig.~\ref{fig3} for the LO mode and Fig.~\ref{fig4} for the TO mode. For LO phonons, three KAs which correspond to the non-analyticities of $\Pi_+^{\mathrm{pp}}$ are evident, occurring at the wavevector $vq = \omega_0$, $vq = 2\varepsilon_F\pm \omega_0$. For TO phonons, the divergence of $\Pi_+^{\mathrm{pp}}$ at $vq = \omega_0$ is removed due to the self-consistency condition for $\omega$ in Eq.~(\ref{eq4}), but the KA remains as a sharp but finite peak at $vq = \omega_0$. In addition, we note the KA at $vq = \omega_0$ for both the LO and TO modes, unlike the other two KAs for the LO mode, is independent of electron density, an interesting consequence of the quasi-relativistic linear dispersion peculiar to graphene. For both LO and TO phonons, our results suggest that the phonon energy first increases (i.e., phonon hardening) with density up to a certain phonon wavevector, and then decreases (i.e., phonon softening) with density. The critical wavevector for this transition from phonon hardening to softening is different for LO and TO phonons, and we find numerically $q \simeq 5\times10^8\mathrm{m}^{-1}$ for LO phonons and $q = \omega_0/v = 3\times10^8\mathrm{m}^{-1}$ (i.e., the KA) for TO phonons. 

In the following, we provide a schematic picture for understanding the occurrence of the KAs. The inset of Fig.~\ref{fig4} shows
the $\omega-q$ phase space in which single-particle excitation can occur through virtual phonon exchange. The behavior of the phonon self-energy is characterized by six different regions of the $\omega-q$ phase space where $\Pi_+^{\mathrm{pp}}$ is analytically continuous, separated by the boundaries (indicated by the solid lines) corresponding to the set of values of $(\omega,q)$ where $\Pi_+^{\mathrm{pp}}$ is non-analytic. \textit{Necessary} (but by no means sufficient) conditions for the KAs to occur are given by the intersection points between the phonon dispersion line $\omega = \omega_0$ and the boundaries for the different regions. Within the BA, phonons are treated as static with $\omega = 0$ in the expression of the phonon self-energy, the only intersection points therefore occur at $q = 0$ and $q = 2k_F$, as in the case of, e.g., a usual metal. In graphene where the BA is invalid, phonon dynamics must be included with $\omega = \omega_0$ in the phonon self-energy, the intersection points now occur at $vq = \omega_0$, $vq = 2\varepsilon_F-\omega_0$, and $vq = 2\varepsilon_F+\omega_0$. For the LO mode, we find that KAs occur at all three intersection points. This however does not apply for the TO mode, and we find two equal but opposite contributions in the expression for $\Pi_{\mathrm{TO}+}^{\mathrm{pp}}$ which cancel the effects of two KAs, yielding only one KA at $vq = \omega_0$ in this case. In ordinary metals, the KA $q = 2k_F$ corresponds to backscattering with maximum phonon wavevector $q$ \cite{KA}; this KA becomes $q = 2k_F\mp \omega_0/v$ in graphene because of the phonon non-adiabaticity, with $q = 2k_F-\omega_0/v, 2k_F+\omega_0/v$ corresponding, respectively, to emission and absorption of a phonon through electron backscattering. The KA $q = \omega_0/v$ does not have an analogue in ordinary metals, and corresponds to forward scattering of the electron through absorption of a phonon.

In conclusion, we have developed a theory for the interaction-induced phonon renormalization in graphene, and discovered new and multiple KAs in the renormalized phonon dispersion. The peculiarity and the distinction of these KAs from the usual KAs in metals are signatures of the renormalized dynamically screened e-ph interaction and the special chiral structure of the e-ph coupling in graphene. The graphene phonon energy dispersion can be measured with double resonance Raman scattering or electron energy loss spectroscopy on a monolayer graphene, and the experimental verification of our predictions would establish that graphene has a very unique e-ph many-body coupling.





This work is supported by US-ONR, NSF-NRI, and SWAN SRC.


\begin{thebibliography}{18}
\bibitem{Rem1}  The BA (also going by the name of static or adiabatic approximation) requires the phonon energy to be much smaller than the characteristic energy scale for the electrons, usually the Fermi energy.
\bibitem{Rem2} In this connection, we note that the BA also breaks down in usual doped semiconductors at low enough densities (i.e., $\varepsilon_F \sim \omega_0$) when plasmon-phonon coupling becomes important.
\end{thebibliography}
\end{document}